\pdfoutput=1
\documentclass{PoS}

\usepackage{amsmath,amssymb}
\usepackage{xspace}
\usepackage{tabularx}
\usepackage{subfigure}
\usepackage[sort&compress,numbers,merge]{natbib}
\usepackage{natbib,natbibspacing}
\newcommand{\msbar}{%
    \ifmmode%
        \overline{\mathrm{MS}}
    \else%
        \(\overline{\mathrm{MS}}\)\xspace
    \fi
}
\newcommand{\ripmom}{%
    \ifmmode \mathrm{RI^\prime\text{-}MOM}
    \else \(\mathrm{RI^\prime\text{-}MOM}\)\xspace%\!
    \fi}
\newcommand{\mom}{%
    \ifmmode%
        \mathrm{MOM}
    \else%
        \(\mathrm{MOM}\)\xspace
    \fi}
\newcommand{\tr}{\operatorname{tr}}

\title{Using NSPT for the Removal of Hypercubic Lattice Artifacts}

\ShortTitle{Removal of Hypercubic Artifacts}

\author{\speaker{Jakob Simeth}\\
       Institut f\"ur Theoretische Physik, Universit\"at Regensburg, 93040 Regensburg, Germany\\
    E-mail: \email{jakob.simeth@physik.uni-regensburg.de}}

\author{Andr\'e Sternbeck\thanks{Former address: Institut f\"ur Theoretische Physik, Universit\"at Regensburg, 93040 Regensburg, Germany.}\\
 Theoretisch-Physikalisches Institut, Friedrich-Schiller-Universit\"at Jena, 
 07743 Jena, Germany}

\author{Meinulf G\"ockeler\\
     Institut f\"ur Theoretische Physik, Universit\"at Regensburg, 93040 Regensburg, Germany}

\author{Holger Perlt\\
 Institut f\"ur Theoretische Physik, Universit\"at Leipzig, 04109 Leipzig, Germany}
\author{Arwed Schiller\\
 Institut f\"ur Theoretische Physik, Universit\"at Leipzig, 04109 Leipzig, Germany}

\abstract{
    The treatment of hypercubic lattice artifacts is essential for the calculation of non-perturbative 
    renormalization constants of RI-MOM schemes. It has been shown that for the RI'-MOM scheme a large 
    part of these artifacts can be calculated and subtracted with the help of diagrammatic Lattice Perturbation Theory (LPT).
    Such calculations are typically restricted to 1-loop order, but one may overcome this limitation and calculate
    hypercubic corrections for any operator and action beyond the 1-loop order using Numerical Stochastic 
    Perturbation Theory (NSPT). In this study, we explore the practicability of such an approach and consider, as a
    first test, the case of Wilson fermion bilinear operators in a quenched theory. Our results allow us to compare
    boosted and unboosted perturbative corrections up to the 3-loop order. 
}

\FullConference{The 32nd International Symposium on Lattice Field Theory,\\
		23-28 June, 2014\\
		Columbia University New York, NY}

\begin{document}

\section{Introduction}
Compute power increases every year and allows us to reduce statistical errors of lattice simulations. 
Systematic errors however cannot be cured as simple as that and may pose challenging problems. 
A well-known unknown is the uncertainty due to the finite lattice spacing, which may have a significant effect. For 
determinations of renormalization factors in the context of RI-MOM schemes 
\cite{Martinelli:1994ty} the finite lattice spacing causes for instance so called hypercubic lattice artifacts 
in the data for the renormalization factors. Their origin lies in the breaking of 
the Euclidean continuum symmetry \(O(4)\) to the hypercubic subgroup of reflections and permutations, \(H(4)\). 
This has the consequence that lattice results for the renormalization factors in momentum space depend on 
the direction of momentum, i.e.~the orbit under the action of \(H(4)\). To leading order in the momentum
this discretization effect typically grows, but their exact functional form depends on the operator, the 
lattice spacing and the momentum. For the renormalization scale $\mu$ one therefore requires \(\Lambda^2 \ll \mu^2\ll 
\frac{1}{a^2}\). \(\Lambda\) is the typical scale where nonperturbative effects are present 
and so the first condition ensures that perturbation theory can be applied, e.g., to connect renormalization 
factors to the $\overline{MS}$ scheme. The second condition ensures discretization effects are small. In practice, 
however, both conditions are not met simultaneously and thus hypercubic artifacts often spoil the accuracy of the 
determination.

There are few ways to alleviate this problem. One is to consider only diagonal momenta \(ap=a(q,q,q,q)\),
where \(q=2\pi k_\mu/N_\mu\),\, \(k_\mu=-N_\mu/2+1,\ldots,N_\mu/2\) and \(N_\mu\) is the number of lattice sites in 
direction \(\mu=1,2,3,4\). In these momentum directions hypercubic artifacts are much smaller than in any other 
direction but as we will see a significant amount still remains.

A more sophisticated way to remove these discretization errors is perturbative subtraction.
There, the idea is (1) to calculate the hypercubic artifacts \(\delta\mathcal{O}(a,p)\) for 
a lattice quantity \(\mathcal{O}_{lat}(a,p)\) within lattice perturbation theory (LPT)
and then (2) to subtract this from the nonperturbative data for \(\mathcal{O}_{lat}\). If this subtraction is complete 
the corrected \(\mathcal{O}_{lat}\) depends on $a^2p^2$ rather than $ap$:
\begin{equation}
    \mathcal{O}_{corr}(a^2p^2) = \mathcal{O}_{lat}(a,p) - \delta \mathcal{O}(a,p)\text{.}
\end{equation}
In practice, such a subtraction has been used in \cite{Gockeler:2010yr} using diagrammatic LPT calculations at 1-loop order 
in \(g_0^2\) and also in \cite{Constantinou:2013ada, Constantinou:2012dt} to order \(a^2\).
However, a 1-loop subtraction often appears to be insufficient.

We therefore explore Numerical Stochastic Perturbation Theory (NSPT) as a tool to calculate higher-loop corrections 
to \(\delta\mathcal{O}(a,p)\). We have successfully pursued this approach already for the gluon and ghost propagators in 
Landau gauge \cite{Simeth:2013ima}. Here we will show that NSPT works equally well for \ripmom 
renormalization factors. 
%As mentioned, these factors are not only affected particularly strong by discretization errors 
%but also needed at high precision, because any error would add to the final error of the 
%corresponding observable.
To make the hypercubic artifacts even stronger for our study we use standard (unimproved) Wilson fermions. It is also 
sufficient to look at quenched QCD to see that our method works. This choice was simply for practical reasons 
and unquenched simulations do not impose any problems for future determinations. Neither does the use 
of improved actions: If one is able to subtract large discretization effects which go along with unimproved Wilson 
fermions, one should have good chances to remove the smaller errors one typically sees for improved actions and 
operators.

\section{\ripmom factors from Numerical Stochastic Perturbation Theory}
Numerical Stochastic Perturbation Theory (NSPT) makes use of the Langevin process to provide expectation values for 
observables in the perturbative regime of lattice QCD. It is reviewed in detail in \cite{DiRenzo:2004ge}, and it 
suffices to mention here that for the Langevin updates and all measurements we treat link 
fields (NSPT configurations) and derived quantities as power expansions in \(\beta^{-1/2}=g_0/\sqrt{2 N_c}\),
\begin{equation}
    U_{x,\mu} = 1 + \sum_{i=1}^{n} \beta^{-i/2} U^{(i)}_{x,\mu} + \mathcal{O}(\beta^{-n/2+1})\,.
\end{equation}
Hence, all algebraic operations are performed with respect to this expansion and so additions and multiplications are 
as for polynomials: \(\left[A+B\right]^{(n)} = A^{(n)} + B^{(n)}\) and \(\left[AB\right]^{(n)} = 
\sum_{i=0}^n A^{(i)}B^{(n-i)}\), respectively. An expansion can be inverted using the recursive formula
\begin{equation}
    \left[A^{-1}\right]^{(n)} = -[A^{(0)}]^{-1} \sum_{i=0}^{n-1} A^{(n-i)} [A^{-1}]^{(i)}\text{.}
    \label{eq:nsptinversion}
\end{equation}
The latter we use for example to calculate two- and threepoint functions from the standard Wilson Dirac operator, 
making use of the well-known zeroth-order inverse, the tree-level Feynman propagator $S_0$.
In our setup, we solve the Langevin equation using the simplest Euler integration scheme for three different
step sizes \(\epsilon=0.01, 0.02, 0.03\) and extrapolate our results afterwards linearly to \(\epsilon=0\).
We simulate lattices of sizes \(N^4=16^4, 24^4\) and \(32^4\) to resolve finite size effects.

Renormalization factors in the \ripmom scheme are defined in the chiral limit, for example, via the \(H(4)\)-invariant 
renormalization condition
\cite{Constantinou:2013ada}
\begin{equation}
    {Z^{\ripmom}_\psi}^{-1}(\mu)\cdot Z^\ripmom(\mu) \frac{\sum_{i=1}^{d} 
\tr\left[\Gamma_i(p)\Gamma_i^{Born\dagger}\right]}{\sum_{j=1}^{d} 
\tr\left[\Gamma_j^{Born}\Gamma_j^{Born\dagger}\right]}\bigg|_{\mu^2=p^2} = 1\text{,}
    \label{eq:rimomrenormcondition}
\end{equation}
where \(Z^\ripmom(\mu)\) is the renormalization factor for a flavor non-sing\-let current in the \ripmom-scheme
and \(Z^{\ripmom}_\psi\) is the wave function renormalization
\begin{equation}
    Z^{\ripmom}_\psi(\mu) = \frac{\tr\left[-i\sum_\mu\gamma_\mu \sin(ap_\mu) a S^{-1}(p)\right]}{12\sum_\nu\sin^2(ap_\nu)}\bigg|_{\mu^2=p^2}\;\text{,}
    \label{eq:rimomzpsi}
\end{equation}
On the lattice, one defines the amputated vertex function, \(\Gamma(p) = S^{-1}(p) G(p) S^{-1}(p)\) using the 
threepoint function \(G(p)\) we get from the expectation value of a local quark current \(\mathcal{O}(x) = 
\bar{u}(x)\Gamma d(x)\), where \(\Gamma\) is an interpolator with the desired quantum numbers:
\begin{equation}
    G(p) = \frac{a^{12}}{V} \sum_{x,y,z}\exp(-ip\cdot(x-y))\left\langle u(x) \bar{u}(z) \Gamma d(z) 
\bar{d}(y)\right\rangle\text{.}
    \label{eq:vertexfunction}
\end{equation}
These expressions are suitable both for the perturbative and the nonperturbative case and both we need here. So we 
proceed in a similar way: As fermionic n-point functions depend on the gauge, we first fix both the NSPT and the 
nonperturbative (quenched lattice QCD) configurations to Landau gauge and measure the two- and threepoint functions. 
The only difference is in the use of the algebraic operations for NSPT and how the chiral limit is achieved: The 
nonperturbative $Z$-factors are obtained on a \(32^4\) lattice from a linear extrapolation to zero quark mass of data 
for three values of the hopping parameter \(\kappa=0.1489, 0.1507\) and \(0.1520\) \cite{Gockeler:1998rw}.
For NSPT we can measure directly in the chiral limit, as we only need the tree-level Feynman propagator
for the inversion of the Dirac operator (cf.~\eqref{eq:nsptinversion}) to obtain all n-point functions.
\section{Hypercubic Corrections}
Our perturbative and nonperturbative lattice configurations were thermalized with respect to the same 
(quenched) action. Data accumulated on either set should therefore show similar hypercubic artifacts. For the 
perturbative data we also know -- up to a constant --  their exact values in the continuum limit. This allows us to construct 
the hypercubic artifacts to a given loop order~\(n\)
\begin{equation}
    \delta Z_{\ripmom}^{(n)}(ap) = Z^{(n)}_{NSPT}(ap) - Z^{(n)}_{exact}(ap)\text{,\quad where }\quad
%\end{equation}
%The exact value has the form
%\begin{equation}
    Z^{(n)}_{exact}(ap) = d_{n,0} + \sum_{i=1}^{n} d_{n,i} \log((ap)^2)^i\text{,}
\label{eq:hypcor}
 %\label{eq:zexact}
\end{equation}
and the \(d_{n,i>0}\) can be calculated from the anomalous dimension of the operator under consideration%
\footnote{See, e.g.,~\cite{Brambilla:2013sua} for a similar calculation.}
so that only the constant \(d_{n,0}\) remains to be fixed from the NSPT data.

In practice, however, the data is still afflicted with finite volume effects and possibly large autocorrelation
times from the Langevin process. 
Thus, we do fits to the form
\begin{equation}
    Z^{(n)}_{NSPT}(ap,pL) = d_{n,0} + \sum_{i=1}^{n}d_{n,i} \log((ap)^2)^i + \frac{c_V}{(apN)^2} +
    \sum_{i=1}^{d}\sum_{j=0}^{i-1} c_{i,j} a^{2(i-j)}\frac{p^{[2i]}}{p^{[2j]}}\text{,}
\end{equation}
where the first term is the finite constant we are ultimately interested in and the second term removes the
logarithmic dependence. The third term parameterizes 
the finite size effects in a similar way as in~\cite{DiRenzo:2014csa} and the last sum is a suitable form to describe the hypercubic artifacts in our fit
range \([(ap)^2_{min}, (ap)^2_{max}]\) up to a degree \(d\), corresponding to a \(\mathcal{O}(a^{2d})\) removal.
The free parameters of the fit are \(d_{n,0}\), \(c_V\) and \(c_{i,j}\) and their number and values depend
on the choice of the fit range and \(d\). The latter needs to be increased with the fit range to describe the 
finite lattice spacing effects.

To estimate the error correctly, and to cover the interesting region of small \( (ap)^2\) while
avoiding too strong systematic effects, we bootstrap over our data points and do various fits
with different values of \((ap)^2_{min}\), \( (ap)^2_{max}\) and \(d\). Finally, we take the weighted mean
of the individual results using the probability density from the corresponding \(\chi^2\) distribution.

Our results for \(d_{n,0}\) can be read off from Table~\ref{tab:res} a) up to three loop order. Table~\ref{tab:res} b) 
compares the 1-loop results with values from the literature. Overall, the values are in good agreement, 
although the given statistical errors still seem to underestimate the true errors.
\begin{table}
    %\centering
%& \(d_{i,0}^\psi\) 
%& \(-0.825(8)\)    
%& \(-1.36(2)\)     
%& \(-3.68(7)\)     
    \vspace*{-0.3cm}
    \hspace*{-0.3cm}
    \resizebox{\textwidth}{!}{
    \subtable[Results for finite contributions\label{tab:finiteconsts}]{
    \begin{tabularx}{0.6\textwidth}{l|c|c|c|c}
        \(i\) & \(d_{i,0}^V\) & \(d_{i,0}^{AV}\) & \(d_{i,0}^S\) & \(d_{i,0}^T\) \\\hline\hline
        \(1\) & \(-1.045(3)\)  & \(-0.808(3)\) & \(-0.862(3)\)  & \(-0.868(3)\) \\ % & \(-0.416(34)\)\\
        \(2\) & \(-1.93(2)\) & \(-1.36(2)\) & \(-1.65(3)\) & \(-1.47(2)\) \\ % & \(-0.929(80)\) \\
        \(3\) & \(-5.82(5)\)   & \(-3.79(3)\)  & \(-5.42(14)\)  & \(-4.06(4)\)%  & \(-2.63(28)\) 
\end{tabularx}}
\subtable[Comparison to LPT\label{tab:lptcmp}]{
    \begin{tabularx}{0.41\textwidth}{l|c|c|l}
        & NSPT & LPT & ref. \\\hline\hline
        %\(Z^{(1)}_\psi\) & \(-0.825(8)\) & \(-0.843216\) & \cite{Skouroupathis:2008mf}\\
        \(Z^{(1)}_V\) & \(-1.045(3)\) & \(-1.044510\) & \cite{Skouroupathis:2008mf}\\
        \(Z^{(1)}_{AV}\) & \(-0.808(2)\) & \(-0.800249\) & \cite{Skouroupathis:2008mf}\\
        \(Z^{(1)}_S\) & \(-0.862(3)\) & \(-0.858819\) & \cite{Skouroupathis:2007jd} \\
        \(Z^{(1)}_T\) & \(-0.868(3)\) & \(-0.862146\) & \cite{Skouroupathis:2008mf}\\
\end{tabularx}}
}
\caption{
    a) Final results for the finite contributions -- i.e. the \(d_{i,0}\) in Eq.~\protect\eqref{eq:hypcor} -- of the renormalization constants for the vector (V), axialvector (AV), scalar (S) and tensor (T) currents.
    b) Comparison between the NSPT results of this work and known 1-loop LPT calculations.
}
\label{tab:res}
\end{table}

With our values for \(d_{n,0}\), we can now quantify the hypercubic discretization errors, using 
Eqs.~\eqref{eq:hypcor}. The left panels of Fig.~\ref{fig:cor_vectorscalar} 
show these corrections summed up for the vector and scalar renormalization factor 
setting \(\beta=6.2\). These corrections can then be used to improve the nonperturbative data thermalized for the same 
$\beta$,
\begin{equation}
    Z_{corr}(a,p) = Z_{NPR}(a,p) - \sum_{i=1}^3 \beta^{-i} \delta Z^{(i)}(a,p)\text{.}
    %might want to include finite size correction?
\end{equation}
Data for $Z_V$ and $Z_S$ before and after this subtraction is shown in the right panels of 
Fig.~\ref{fig:cor_vectorscalar}.
\section{Boosted Perturbation Theory}
Tadpole contributions to perturbative expansion coefficients 
are often large in lattice perturbation theory. Such diagrams do not have a continuum analogue and
are the main reason for the slow convergence of LPT \cite{Lepage:1992xa}. Of course, also the NSPT results obtained here
and the subtraction of discretization errors suffer from this problem.

The solution is to rescale the expansion parameter by the plaquette \cite{Lepage:1992xa}. 
The rationale for that is that tadpole diagrams always contain a purely gluonic loop
which is proportional to the plaquette. 
Hence, one defines a ``boosted'' or ``tadpole-improved'' coupling 
\(g_0^2 \to g^2_b = \frac{g^2_0}{P}\),
or equivalently,
\(\beta \to \beta_b = P\,\beta\),
where \(P=1+\sum_{i=1}\beta^{-i}p^{(i)}\) is measured perturbatively but in the usual way.
Both the unboosted and boosted series of a quantity \(\mathcal{O}\) with coefficients \(o_0^{(i)}\) and \(o_b^{(i)}\), 
respectively, 
should converge to the same value:
\begin{equation}
    \sum_{i=0}^\infty \beta^{-i} o_0^{(i)} = \sum_{i=0}^\infty \beta_b^{-i} o_b^{(i)}\text{.}
\end{equation}
By inserting \(\beta^{-1} = P\,\beta_b^{-1}\), we express the unimproved coupling in terms of the new, shifting
the dependence on the old expansion parameter
one order higher. To 3-loop order we find by comparing the coefficients
of \(\beta_b^{-1}\)
\begin{equation}
    o_b^{(0)} = o_0^{(0)},\,
    o_b^{(1)} = o_0^{(1)},\,
    o_b^{(2)} = o_0^{(2)} + p^{(1)} o_0^{(1)},\,
    o_b^{(3)} = o_0^{(3)} + 2\,p^{(1)}o_0^{(2)} + p^{(2)} o_0^{(1)} + (p^{(1)})^2 o_0^{(1)},\,
     \cdots \text{,}
     \label{eq:boostedcoeffs}
\end{equation}
so that one obtains the improved \(n\)-loop estimate 
\(\mathcal{O} \approx \sum_{i=0}^{n} \beta_b^{-i} o_b^{(i)}\),
which is, for any finite truncation at order \(n\), different from the original expansion.

\begin{figure}
    \vspace{-1.0cm}
    \resizebox{0.5\textwidth}{!}{
        \includegraphics{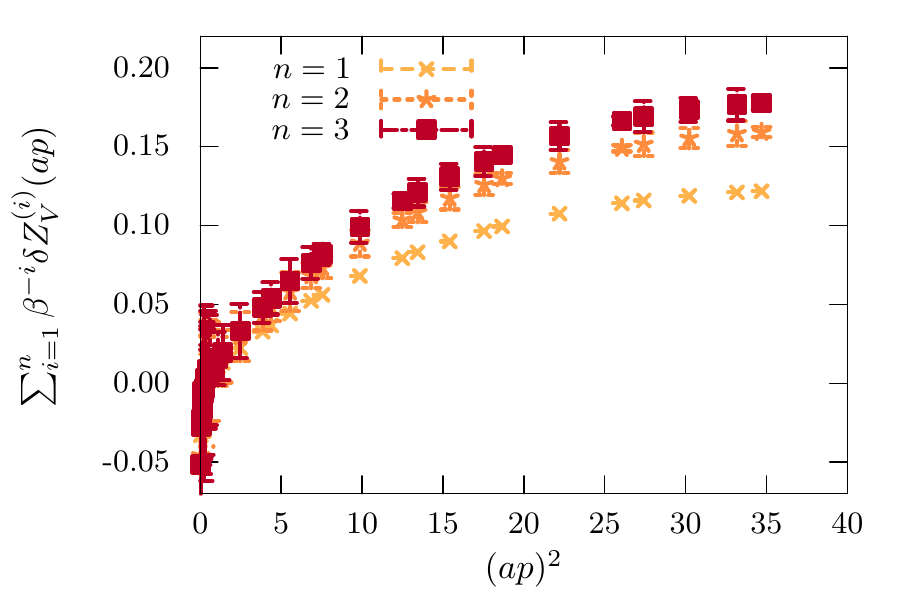}
    }
    \resizebox{0.5\textwidth}{!}{
        \includegraphics{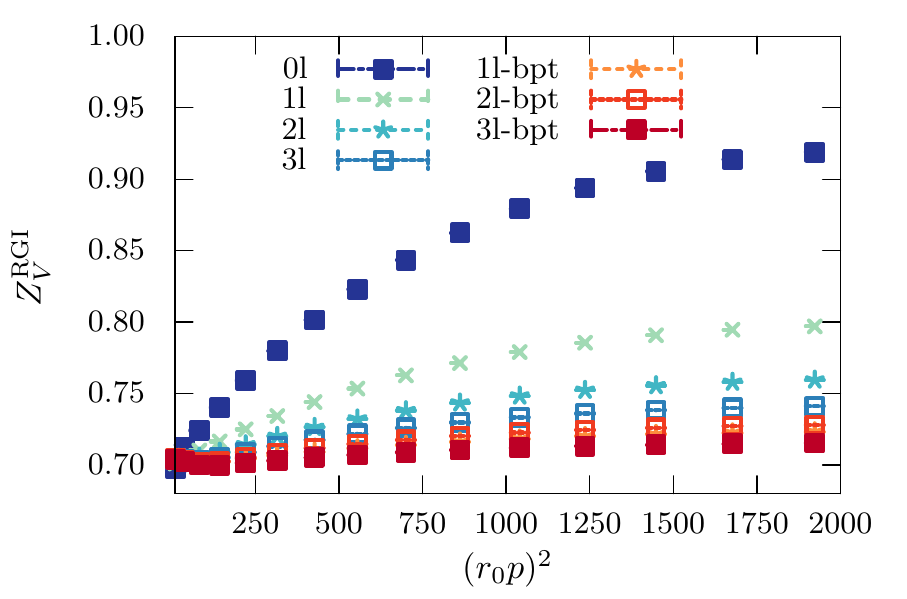}
    }
    \resizebox{0.5\textwidth}{!}{
        \includegraphics{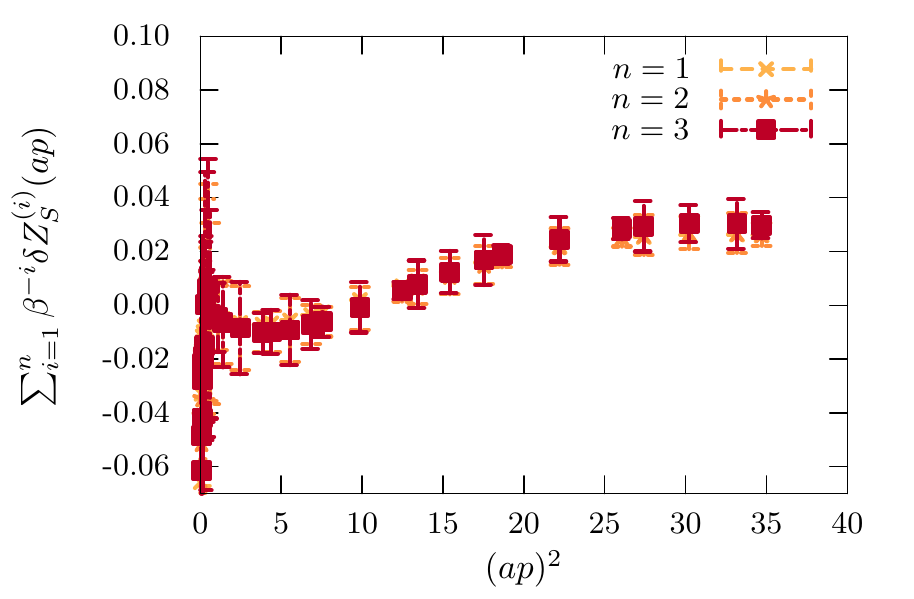}
    }
    \resizebox{0.5\textwidth}{!}{
        \includegraphics{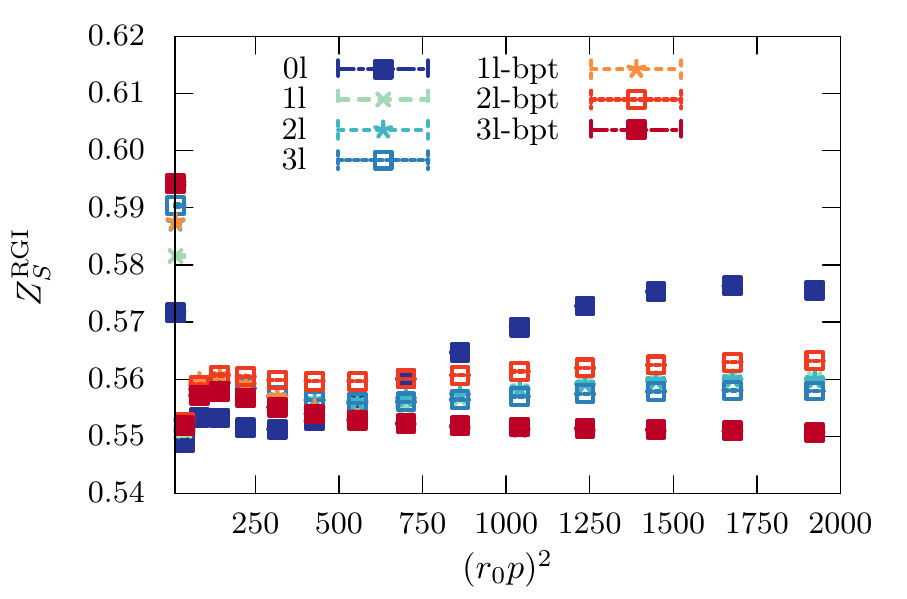}
    }
    \caption{(Unboosted) Hypercubic corrections (left) and resulting plateaus for the unimproved (0l), unboosted improved (1l to 3l) and boosted (1l-bpt to 3l-bpt) RGI vector (top) and scalar (bottom) renormalization factor.}
    \label{fig:cor_vectorscalar}
\end{figure}

In our case, \(\beta=6.2\) gives \(P=0.623\), so that \(\beta_b = 3.864\).
It turns out that boosted perturbation theory is very efficient in all cases: The boosted 1-loop subtraction has already
a larger effect than the unboosted 3-loop subtraction.
The right-hand sides of Fig.~\ref{fig:cor_vectorscalar} show the
subtractions to different loop-orders and compare the boosted with the unboosted results for the examples of the
scalar and the vector renormalization factor.

\begin{figure}
    \resizebox{0.5\textwidth}{!}{
        \includegraphics{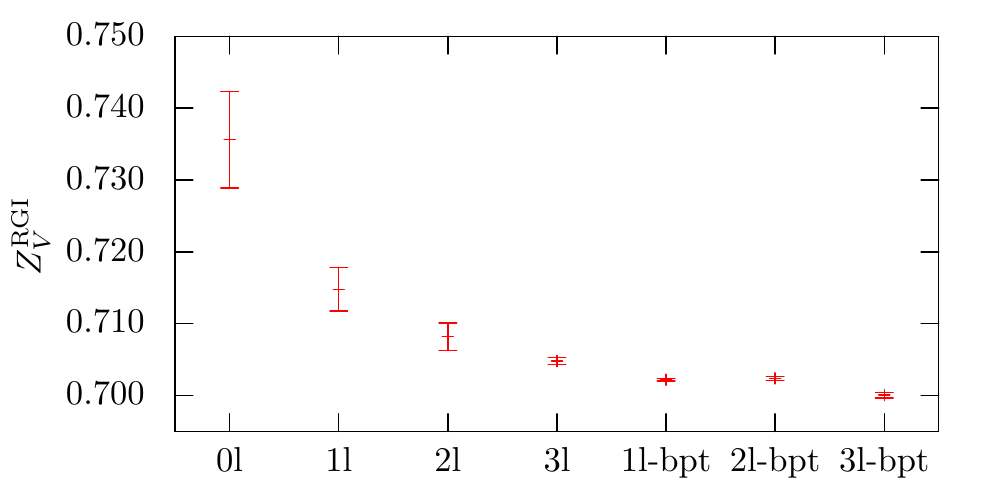}
    }
    \resizebox{0.5\textwidth}{!}{
            \includegraphics{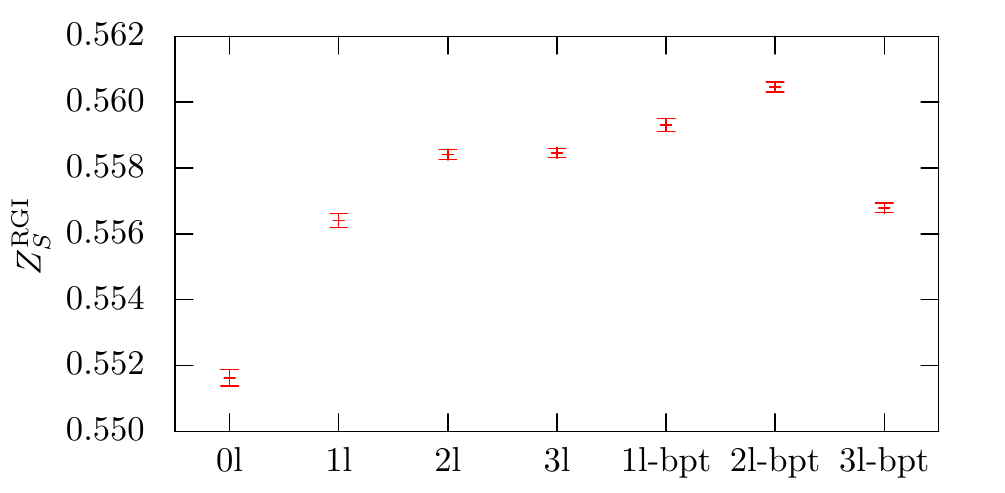}
        }
    \caption{Fits of the Vector (left) and Scalar (right) RGI renormalization factor to a constant. The fit range 
     was fixed to \(a^2p^2 \in [2,12]\). The abscissa labels the degree of data improvement: 0l = unimproved,
     nl(-bpt) = improved by our corrections up to n-loop order in the bare (boosted) 
     coupling. Errors are purely statistical.}
    \label{fig:fitcmp}
\end{figure}

Boosted Perturbation Theory (BPT) has already been
applied in the subtraction of 1-loop discretization effects \cite{Gockeler:2010yr} and, as it turns out now,
this is a very successful and efficient way to remove most of the artifacts.
Nevertheless, also in boosted perturbation theory, the 3-loop subtraction gives in most cases a significant 
improvement over the lower orders. A pretty stable plateau is reached over nearly the whole range of momenta.
The scalar renormalization factor seems to be an exception to the rule and raises slight doubts on the applicability 
of boosted perturbation theory in certain cases: The boosted 1-loop curve lies on top of the boosted 3-loop curve,
whereas the 2-loop subtracted result lies even above the unboosted 1-loop subtraction.
This can also be seen from Fig.~\ref{fig:fitcmp} which shows fits to a smaller plateau region for the two examples, using different methods of improvement.
The reason lies in the smallness of the (unboosted) 2- and 3-loop coefficients of the correction,
so that the boosted 2- and 3-loop coefficients are dominated by the (negative) plaquette contributions 
(cf. Eq.~\eqref{eq:boostedcoeffs}). This has the consequence that the boosted coefficients alternate in sign
and the boosted subtraction probably alternates around the true value. Thus any finite subtraction might well ``overshoot'' the true value.
%\begin{figure}
%    \include{fits/cmp_av}
%\end{figure}
%\begin{figure}
%    \include{fits/hypcor_zAV}
%    \include{fits/bpt-cor-Axialvector-rgi}
%\end{figure}
\section{Conclusions}

Using NSPT we were able to calculate the leading hypercubic lattice corrections of several hadronic operators 
in quenched (boosted) lattice perturbation theory (LPT) with $N_f=2$ valence Wilson 
fermions. Our results serve up to 3-loop order in an expansion of 
the gauge coupling $g^2_0$ (resp.\ $g_b^2$) and allow us to remove almost all hypercubic lattice artifacts in the 
corresponding \emph{nonperturbative} data for the renormalization factors $Z_V$, $Z_{AV}$, $Z_S$ and $Z_T$ in the RGI 
scheme.

Looking at these corrections in more detail we find that a calculation at 3-loop order in $g_0^2$ provides 
about the same corrections as a 1-loop calculation in $g_b^2$. Moreover, in contrast to $g_0^2$, the additional 
corrections from a 2- or 3-loop calculation in $g_b^2$ would provide only a negligible further improvement for 
$Z_V$, $Z_{AV}$, $Z_S$ and $Z_T$ (see Fig.~\ref{fig:fitcmp}), if diagonal lattice momenta are considered. 
This might be different for operators with derivatives and for off-diagonal momenta, which will be 
further analyzed in a forthcoming publication. Note also that boosted LPT leads to alternating contributions for $Z_S$ 
for different loop orders.

To provide a proof of principle for our approach, we chose a partially quenched setup with unimproved Wilson 
fermions, but we are not restricted to that. We have demonstrated that with NSPT one can estimate the momentum 
dependence for the leading hypercubic lattice corrections of different hadronic operators
beyond the 1-loop order. Such calculations are indeed challenging but not as they were within 
traditional LPT, in particular with respect to all the improved actions used in state-of-the-art 
lattice QCD simulations.

\bigskip
%\acknowledgments
{\small
This work was supported by the European Union under the Grant
Agreement IRG 256594 and the SFB/TRR-55 ``Hadron Physics from Lattice QCD'' by the DFG.
H.P. was supported by Deutsche Forschungsgemeinschaft, DFG Grant: SCHI 422/9-1.
Computation time on the Linux-Cluster of the Leibniz
Rechenzentrum in Munich and the iDataCool in Regensburg are acknowledged.}
%---------------------------------------------------------------------------
%
\bibliographystyle{apsrev4-1}
{\small \bibliography{references}}
\end{document}